# The Role of Marketing in Public Policy Decision Making: The Case of Fuel Subsidy Removal in Nigeria

Ighomereho, O. Salome[1], Ezeabasili E. Ifeoma[2]
[1](Department of Economics and Business Studies, Redeemer's University, Ogun State, Nigeria)
[2](Department of Political Science and Public Administration, Redeemer's University, Ogun State, Nigeria)

***Abstract:*** *Public policy decision making has become more complex and complicated in recent times. Some authors have attributed this to the fact that public policy decision makers now have more variables to consider in every decision more than ever before. Others have argued that the rate of civilization, globalization and information technology has made the public to be more enlightened and abreast with the activities of government and so can oppose government decisions if they are unfavourable. This tends to increase government need for more and better information in order to satisfy the public. Consequently, this paper examined the issue of fuel subsidy removal in Nigeria, the impact of the policy on the public as well as the country and the role marketing principles would have played if the Nigerian government had taken some time to investigate what should be done, how it should be done and when it should be done. It also proposed a roadmap for future policies that have direct implications for the general public.*

## I. Introduction

Government at all levels makes numerous strategic and tactical decisions in the process of identifying and rendering services to the public. These decisions are complicated by environmental factors such as general economic conditions, laws, political environment, social and cultural changes. According to Bridgetown (2007) public problems are complex and ill-defined and that their consequences may not be identifiable before implementation. These factors have resulted to conflict of perception between the government and the public in most public decisions in Nigeria. Policy makers however, are expected to act on the basis of the perceptions of public interest or of beliefs about what is appropriate or ethically correct in public policy. This could be achieved with the aid of marketing principles and concepts. Marketing is all about identifying the needs and wants of customers (public) as well as other stakeholders and making every effort to satisfy them (Kotler and Keller, 2006). Gundlach (2007) as well as Lusch (2007) have argued that marketing can be very useful in different areas of human life and society. Marketing can serve not only business organizations but also the goals of society. In political marketing, the public could be referred to as the customers whom the government intends to satisfy. Hogan (2005) emphasized the need for accurate, timely data and analysis concerning the state, its communities, its citizens, and the economy for government to carry out effective policy decision making. Such data and analysis help to inform the government about policy alternatives, to monitor progress towards strategic goals, and to help in managing public activities. Marketing research can be very useful for such purposes. It is conceptualized as the systematic and objective identification, collection, analysis, and dissemination of information for the purpose of assisting marketers (government) in making effective decisions relating to the identification and solution of problems and opportunities. According to (Beri, 2008) it specifies the information required to address issues, designs the method for collecting information, manages and implements the data collection process, analyzes the results, and communicates the findings and their implications.

However, as noted over three decades ago by Wilkie and Gardner (1974) and Dyer and Shimp (1977) and is still evident today, "there is underutilization of marketing research in public policy decision making". Very seldom does the policy maker have the time to conduct an extensive research investigation on what the public (customers) actually want, how they want it and when they want it. Where it is conducted, it is deliberately biased so as to result in predetermined findings. Marketing research however, is objective and should be carried out with ethical considerations (Cooper and Schindler, 2006).

On 1st January, 2012, the federal government of Nigeria came up with the policy of fuel subsidy removal. It was argued that the move became necessary since there was no provision for fuel subsidy in the 2012 budget (Alexander, 2012). This decision by the federal government did not go down well with the majority of Nigerians. It led to a week nationwide strike. The strike that started on Monday 9th, 2012 happened to be the most devastating strike action in the history of Nigeria. Every sector of the economy was crippled. Moreover, the Central Bank Governor, Lamido Sanusi put the loss incurred during the strike at $617million daily, translating into about N100 billion (Akanbi and Agbo, 2012). Given this scenario, this paper examined critically

the policy of fuel subsidy removal in Nigeria and justified the difference marketing principles and concepts would have made to the effective design and implementation of the policy. The paper also provided a roadmap for future public policy decision making especially for decisions that have implications for the general public.

## II. The Policy of Fuel Subsidy Removal in Nigeria

Subsidy exists when consumers are assisted by the government to pay less than the prevailing market price of a given commodity. CPPA (2012) defined subsidy as any measure that keeps prices consumers pay for a good or product below market levels for consumers or for producers above market level. In respect of fuel subsidy, it means that consumers would pay below the market price per litre of petroleum product (Ovaga, 2012). Many governments across the globe perceive the provision of subsidies as a social obligation to the economically disadvantaged citizens, particularly the poor and vulnerable groups. The federal government of Nigeria came up with the policy with the aim of reducing the price of the product, thereby minimizing the direct burden on the masses especially the poor. The idea was deemed a good one as it was intended to give the average Nigerian access to cheap petroleum products (National Impact, 2012, Alexander, 2012). However, this has not been the case in Nigeria and in other developing countries. Balouga (2012) reported a study on twenty developing countries (excluding Nigeria) which indicated that although the poorest people benefit a little from the subsidies, the bulk of the benefits go directly to the richest 10 percent. In the sample of the countries in the study, only 7.1 percent of the subsidy benefits go to the poorest 10 percent of the population while the top 10%, on the other hand, gets 47.6 percent of the benefits. It was also found that the top 20% of the population gets about 67.5% of the total subsidies. This shows that in most countries, only a few individuals who are usually rich actually enjoy the benefits of subsidies because it is usually abused.

In Nigeria, the story is not different. As a result of the abuse of fuel subsidy, it became a huge burden on the federal government because the costs of fuel importation continue to increase yearly. As Alexander (2012) observed, it is the total cost of importing fuel into Nigeria that the government and its advocates describe as fuel subsidy. He noted that the pricing template of the Petroleum Product Pricing Regulatory Agency (PPPRA) showed that currently the landing cost of a litre of petrol is N129.21. The margin for transporters and marketers is N15.49, bringing the expected pump price of petrol to N144.70. However, the initial official pump price for petrol was N65 per litre, a difference of N79.70, which the government subsidizes. Recognizing the fact that there are significant investment opportunities in Nigeria's downstream sector if well managed, the focus of the current government is to fully deregulate the sector through the licensing of private refineries, the privatization of the existing ones and the removal of subsidies. Taking a cue from other countries that have privatized, the Nigerian government went ahead with the policy even against the backdrop of widespread disapproval on the part of the public (Balouga, 2012).

Subsidy removal in Nigeria dates back to 1978 when the then military government of Gen. Olusegun Obasanjo reviewed upward the pump price of fuel which was at 9 kobo to 15.37 kobo. The concern was for government to generate enough money to run the administration particularly when it was preparing for the 1979 democratic elections and also to carter for the social needs of Nigerians. Subsequently, fuel subsidy removal became a major policy of every government in Nigeria. At every point in time that the pump price of fuel was increased, the public have always opposed it vehemently. An analysis of the sequence of fuel subsidy removal indicates that only the government of Buhari/Idiagbon and Umaru Shehu Yardua spared Nigerians the ordeal of fuel price increase. This some have argued is because of their focus on fighting corruption and indiscipline in the Nigerian society. The table below indicates the different pump prices by the different administrations from 1978 to 2012:

### III. Table 1: Fuel Price Changes in Nigeria from 1978 to 2012

| S/N | DATE | ADMINISTRATION | PRICE | SUBSIDY WITHDRAWN |
|---|---|---|---|---|
| 1 | 1976 | General Murtala Mohammed | 9k | - |
| 2 | 1978 | Gen. Olusegun Obasanjo (as military ruler) | 15.37k | 6.37k |
| 3 | 1982 | Alhaji Shehu Shagari | 20k | 4.63k |
| 4 | 1986 | Gen. Ibrahim Babangida | 39.50k | 19.50k |
| 5 | 1988 | Gen. Ibrahim Babangida | 42k | 2.50k |
| 6 | 1989 | Gen. Ibrahim Babangida | 60k | 18k |
| 7 | 1991 | Gen. Ibrahim Babangida | 70k | 10k |
| 8 | 1993 | Gen. Ibrahim Babangida | N3.25k | N2.55 |
| 9 | 1994 | Gen. Sani Abacha | N11.00 | N7.75 |
| 10 | 1999 | Abdulsalami Abubakar | N20.00 | N9.00 |
| 11 | 2000 | Olusegun Obasanjo (As a civilian president) | N22.00 | N2.00 |
| 12 | 2002 | Olusegun Obasanjo | N26.00 | N4.00 |

| 13 | 2003 | Olusegun Obasanjo | N40.00 | N14.00 |
| 14 | 2004 | Olusegun Obasanjo | N45.00 | N5.00 |
| 15 | 2007 | Olusegun Obasanjo | N70.00 | N25.00 |
| 16 | 2007 | Alhaji Umaru Shehu Yardua | N65.00 | (N5.00) |
| 17 | 2012 | Dr. Goodluck Jonathan | N97.00 | N32.00 |

**Source:** (Adapted from Ering and Akpan, 2012:15)

In January 1982, the civilian regime of Alhaji Shehu Shagari then raised the pump price to 20 kobo from 15.37 kobo. Money realized from the fuel increase was not used to put in place social services that Nigerians badly needed at that time. The corruption that bedeviled the administration led to its overthrow. General Babangida came in and also increased the pump price of fuel to 39.50 kobo in March 31st, 1986. This regime was notorious for numerous pump price increases. On April 10th, 1988, the regime increased it to 42 kobo from 39.50 kobo per litre and then again to 60 kobo on January 1st, 1989. This decision led to massive protests by Nigerians. The economic down turn coupled with the increases made life really unbearable and Nigerians reacted angrily. Again, on the 6th of March, 1991, the Babangida administration raised the pump price from 60 kobo to 70 kobo. Not too long, in November 8, 1993, the pump price was raised to N5.00. This astronomical increase led to mass protests across the length and breath of Nigeria and so, the price was reduced to N3.25 on November 22, 1993. A year later, on October 2nd, 1994, it was again raised to N15.00 only to be reduced two days later to N11.00 by the Gen. Abacha's regime. The reduction was as a result of mass protests, coupled with the need to win the support of Nigerians. On December 20, 1998, the pump price was also increased to N25 but again reduced to N20 on January 6th, 1999 just a month later. This was during Gen. Abdulsalam Abubakar brief transitional reign as a military ruler. The decision witnessed sustained protests by Nigerians, the organized labour and the Civil Society Organizations (Ering and Akpan, 2012)

When Gen. Olusegun Obasanjo became a civilian president for eight years, Nigeria witnessed several rounds of fuel price increases. The first started on June 1st, 2000, where the petrol price per litre was raised to N30.00 but only to be reduced to N25 one week after due to massive protests by organized labour, civil society organizations and the ordinary Nigerians. Five days later, on June 13, 2000, the pump price was further adjusted to N22.00 per litre. On January 1st, 2002, Obasanjo regime increased the price from N22.00 to N26.00 and to N40.00 on June 23, 2003 just one year after. In 2004, it was again increased to N45. In 2007, the same regime also raised the price of fuel per litre to N70, but the Yaradua's regime later reviewed it downward to N65 on assumption of office in May 2007. This was how it remained until President Goodluck Jonathan regime decided to finally remove fuel subsidy (Ering and Akpan, 2012). The issue of fuel subsidy removal has been on ground since he became president after the April, 2011 general election. However, while the nation wide consultations and discussion on fuel subsidy removal was still going on, the Petroleum Product Pricing Regulatory Agency (PPPRA) on January 1st, 2012, announced the outright removal of fuel subsidy and fixed the pump price of fuel at N141. This led to provocation of Nigerians into action and a face off between the government and the people. Both sides refused to back down for the other, each believing that his views and opinion are better than the other (National Impact, 2012).

This resulted in massive strike actions and protests by the Nigerian Labour Congress (NLC), Trade Union Congress of Nigeria (TUC), Petroleum and Natural Gas Workers Senior Staff Association (PENGASAN), Civil Society Organizations, Academic Staff Union of Universities (ASUU) and the generality of Nigerians. The mass protests almost transformed into the "Nigerian spring" which would have brought down the regime. The regime quickly entered into negotiation with organized labour and rescinded its decision of an outright removal to a partial removal and reduced the pump price to N97.

### IV. Is the Policy the Solution to Nigeria's Challenges?

The controversial issue of fuel subsidy removal has given rise to contending arguments on the merits and demerits of the policy. The federal government maintained that it will help eliminate incentives for corruption and excess profiteering by an unpatriotic cabal in the petroleum sub sector (National Impact, 2012). The removal of fuel subsidy is assumed to affect corrupt politicians more than the masses because their fraudulent practices can no longer be accommodated. The federal government also stated that by removing fuel subsidy, it would be saving about N1.3 trillion per annum (Balouga, 2012), which will free more funds for local investment in the oil sector and increase local refinery production. The fund will also be diversified to improve the lot of a greater majority of the less-privileged Nigerians through the provision of infrastructures that would provide employment for the teeming jobless citizenry as well as improve education, health, power, water resources and agriculture (CPPA, 2012; Nwadialo, 2012).

It is also argued that the removal of fuel subsidy will encourage foreign investment in downstream infrastructure (Onyishi et al, 2012). It will open up the petroleum sector to local and foreign investments just like what happened in the telecommunication sector, which in-turn will create more jobs for Nigerians directly and indirectly. Most importantly, the liberalization of the petroleum sector will reduce government interference, and allow the federal government to focus more on the business of governance, improving social services delivery and building infrastructures. It was noted however that deregulation and removal of the subsidy may initially lead to inflationary pressures but as the market is opened up to investors, billions of dollars will flow into the down stream sector and more private refineries will open for business in Nigeria. Eventually, the market will self regulate and prices for refine petroleum products and other goods and services will be at the market level as competition will tend to force prices down. It is assumed that the long term benefit will be more than the short term pain.

Furthermore, it was argued that it will minimize borrowing and save money for investment in job creation, power and transport infrastructure and others. It will eliminate capital flight and build Nigeria's foreign reserve in order to position the economy for speedy growth and global competiveness. The liberalization and deregulation of the down stream sector of the petroleum industry would finally actualize the objective of ending perennial fuel scarcity and maintenance of sustainable fuel supply across the country (Onyishi et al, 2012). This is because it will trigger private sector investment in a deregulated downstream petroleum sector and enthrone efficiency and catapult the development of the nation's productive sector such as agriculture and industries.

According to the 2012 – 2012 Medium Term Fiscal Framework (MTFF) and the Fiscal Strategy Paper (FSP) which President Jonathan sent to the National Assembly, he stated among other things that fuel subsidy will free up to about N1.3 trillion, that is, about $8 billion dollars in savings. This money he added will be deployed into providing safety nets for various segments of the society which will help to ameliorate the effects of subsidy removal. Also, subsidy removal and the money generated will address the great imbalance between recurrent and capital expenditure in Nigeria and hence, would guarantee the success of the (MTFF). Money realized will be used to build more refineries and buy buses that will help cushion the effect of the subsidy removal.

The justification for subsidy removal by the federal government prompted some Nigerians to argue that the policy of fuel subsidy removal is a good intention but it was poorly executed. The concern is that government should have fixed or put in place a number of measures and infrastructures before going ahead to remove the subsidy. The problem of power should have been fixed so that Nigerians would have to contend only with the fueling of their cars instead of also looking for ways to power their houses, offices and industrial generating plants. New refineries should have been built and the older ones put into functioning so that the availability of the product locally will impact on the economy and play a role in bringing down the price of the product. Others however, have argued that it is not the solution to Nigeria's economic challenges. The Organized Private Sector (OPS) described the policy as a deliberate move by the federal government to worsen the decaying industrial sector. They also argued that it will force them to pay more for providing generating plants at their factories. Similarly, the Small and Medium Enterprises (SMEs) will be seriously affected since most of them use petrol for their relatively smaller power generating plants. As such, it can only impose more hardship on the already economically downtrodden masses. It will cripple further a seriously ill economy. It will have a negative impact on the fragile financial system and send shocks throughout the economy (Okonmah, 2012).

The general perception in the public sphere is that the removal of fuel subsidy has not significantly improved their lives in any way. The policy, which started in 1978, has only benefited successive rulers in Nigeria and the so called "cabals" parading themselves as contractors and multinational companies. The cabal is a group of economic saboteurs who are beneficiaries of the fuel subsidy. The members are known but because they occupy strategic positions in the affairs of this country, they appear to be untouchable. In addition, the revenue from the oil sector has continued to skyrocket such that the price of a barrel of Nigeria's crude oil in the international market stood above 106 dollars as at August 2012. But to what extent have the huge revenues being realized from the sale of approximately 2.3 million barrels daily, impacted on the lives of the poor masses in Nigeria? It is disheartening to observe that Nigeria is the only country that produces oil and has the highest poverty rate in the whole world (Ovaga, 2012). They are of the opinion that since the government has refused to provide basic welfare amenities such as food, good water, good roads, good medical facilities and good transport system then, the government has no justifiable reason why fuel subsidy which is the major resource that has a direct effect on the general public and the economy of the country should be removed

Moreover, the public argued that Nigeria is rated as Africa's second largest producer of crude oil after Libya, and the sixth largest oil producing/exporting country in the world (Nwadialo, 2012). As such, many Nigerians regard cheap fuel as the only benefit they could get from the nation's oil wealth. However, the positive impact of the huge revenues that are realized since the inception of production and exportation of oil in the country has not been felt by most Nigerians (Balouga, 2012). This is because the proceeds from the removal have never been translated into the development of basic infrastructures and social amenities to be enjoyed by the masses. According to Ovaga (2012) Nigeria has the highest pump price per litre of fuel among the OPEC members which include Iran, Kuwait, Qatar, Saudi Arabia, UAE, Venezuela and Libya so, why the increase? A large segment of the population further argued that fuel subsidy removal may not be the only way to save money for the country. Most importantly, they are of the view that the poor masses who already live in abject poverty and can hardly afford the basic necessities of life should not be the ones to bear the cost of transformation while the corrupt leaders who are actually the reason we are at this present condition will not bear any cost. They expressed bitterly that they are not interested in long term gain but are more concern with the level of poverty in the country at present. Some are of the opinion that if the government wishes to be fair, it should bring the cabals to book first before deregulation or preferably they should return stolen public fund to enable the government carry out its transformation agenda. Also, they are not so sure if the savings from subsidy removal will go directly into rehabilitating the refineries as promised. Nzeagwu and Gyamfi (2011) noted that the National Association of Business Students (NANS) stated that "any attempt by the federal government to increase the price of petroleum products in the country by any guise would be met with stiff opposition by Nigeria students and the masses". This is based on the belief that rather than contemplating on increasing the price of petroleum products disguised as removal of subsidy, government should plug the various loopholes that encourage and sustain corruption in the nation's oil and gas sector.

Consequently, there are good reasons to doubt that subsidy removal is the right option as the cabal may only regroup to change tactics, a fact Nigerians are only too aware of. Also, people are skeptical because of the secrecy surrounding the actual amount of the subsidy and how it is being administered by the Nigerian National Petroleum Corporation (NNPC). This goes to prove that the problem of Nigeria does not rely on the removal of oil subsidy but how well funds are managed (Ovaga, 2012). He further stated that the issue of fuel subsidy removal has never been translated into building of refineries and development of basic infrastructure that could ordinarily add value to the lives of Nigerians. That is why many Nigerians remain skeptical about removal of oil subsidy. For over 17 times now, past administrations had removed subsidy in the down stream oil sector at one time or the other, but till today, there is nothing to show for it. The general belief among many Nigerians is that it is corruption that should be removed and not fuel subsidy (National Impact, 2012). The federal government has accused "a cabal" in the petroleum industry to be responsible for the mismanagement of oil subsidy and that most of the products, subsidy beneficiaries claimed to have imported found their way to neighbouring countries through unscrupulous marketers and did not come into the country. The question here is "who are these cabals? And what has been done to them? Although the fuel subsidy probe by the National Assembly has revealed some of the cabals, justice has not been fully implemented. Under these circumstances, the crisis of fuel subsidy in Nigeria could best be resolved through total elimination of corruption in the oil sector, building of new refineries and revamping the existing ones. If these issues are properly addressed, it will go along way to reduce if not bring to a halt the import racketing in the country. It will equally add value to the country's crude oil resources and at the same time minimize the foreign exchange spent on fuel importation (Ovaga, 2012).

## V. The Impact of the Policy on the Public

Some experts, Senators and labour leaders have argued that the federal government did not fully examine the implications of the policy before implementation and that the government will not have enough time or resources to put into effect counter measures to soften the impact of the policy on the average Nigerians. This means that the masses have to bear the brunt because, fuel subsidy removal automatically lead to increases in the pump price of fuel. This was shown by the different pump prices witnessed across the country when the subsidy removal was announced and these ranged from N141 to N200 naira per litre (Alexander, 2012). In some other states of the country, a litre of petrol was sold for as much as N250 naira. Other marketers created artificial scarcity in order to raise the pump price. That is why it is generally believed that outright removal of fuel subsidy will worsen the sufferings of Nigerians.

Many Nigerians argue that Inflation or hyper inflation will ensue as higher energy prices are factored into prices for everything (National Impact, 2012). Basically, the cost of living for the average Nigerian will increase by up to 100% if the subsidy is removed fully, which is inevitable in a country suffering from chronic energy and power shortages, high unemployment and poor infrastructure. In addition, it will automatically increase aggregate costs of production as well as the cost of raw materials and their transportation. With these attendant consequences on small business units, they will probably close down. Although the pump price of petrol was later reduced from N141 to N97 naira, the costs of transportation as well as other products and services did not reflect the reduction. And since the policy was announced during the Christmas period when many Nigerians and their families had traveled to celebrate the Christmas with their families including extended families, many were stranded. Those who could afford it did so by abandoning their families in their villages.

The prices of food stuff also went up drastically. This is because food sellers use transportation to bring in food items and vehicles owners have to struggle to get fuel at very exorbitant prices. The result was that the food sellers had to factor in the increment in order to make marginal gain. School fees and charges were not spared, as school fees have increased. Most parents were left with no choice than to withdraw their children and wards from private schools to public schools where they have to pay less but subject their children/wards to poor education. House rents across the country have increased dramatically and the argument is that fuel price increase has affected the prices of basic building materials and their haulage such as iron rods, roofing sheets, flouring materials and others.

Another impact of the policy is that it led to rebellion against government and anarchy. This was exampled by the massive protests that took place in major cities across the country, after January, 1st, 2012 announcement by Petroleum Product Pricing Regulatory Authority (PPPRA). The removal of fuel subsidy has had a number of negative socio-economic consequences on the Nigerian populace. Even when the federal government has promised and taken a number of rushed and unsustainable remedial measures (palliative) to cushion the effects of the policy on Nigerians, they still were not convinced. The argument is that the so called palliatives should have been put in place before the removal of the subsidy. Even then, the effects of the palliatives are not being felt.

There is also the psychological effect of fuel subsidy removal. It could lead to cases of depression and suicide. It could also result in broken homes and increase cases of divorce. When people cannot fend for themselves and their families, there is the likelihood that husbands and wives would separate. This may consequently lead to discomfort, anger and even death. Due to the negative impact of the policy identified above on the masses, there is need for research into the issues surrounding the policy which will also serve as a guide for future policies that have direct implications for the general public.

## VI. The Role of Marketing in Public Policy Decisions

From the argument on the merits and demerits of the policy, the two schools of thought appear to be convincing. Both sides have good and reasonable argument. However, from the views of the general public on fuel subsidy removal discussed above, it would appear that many Nigerians are not opposed to the policy if they are sure that the funds would be properly accounted for and some issues are addressed. This indicates that the problem really is not the policy but the inability of the federal government to find out what the public really want, how they want it and when they want it. So, there is a missing link between the policy makers and the public. It has been observed that in most developing countries, policies are often driven more by political considerations than by research and rationality (CPPA, 2012) as well as public satisfaction.

According to Bridgetown (2007) there are three decision making frameworks that determine public decisions. They include rational, incremental and political framework. The rational model assumes that the system is stable, the government is a rational and unitary actor and that its actions are perceived as rational choices, well defined objectives are established, alternatives and consequences are known, preferences are clear and there are no limitations of time or cost. In addition, the policy has maximum social advantage that also maximizes the economic advantages. The incremental model assumes that public policy deals with moving targets, the process is not completely rational, analysis is limited, information is ambiguous and subject to interpretation and that different stakeholders may hold varied opinions about means and ends. The process of mutual adjustment of many actors with their own interests and the perceptions about what the public interest is help to reduce or eradicate conflicts and build legitimacy. This model is of the view that a good policy is one over which there is agreement between the policy maker and the public. On the other hand, the political model assumes that policies are political activities and that there are many conflicting conceptions about what the problem is and what the goals should be. Problems are portrayed by actors in a way that promote their favoured course of action, and so, they try to win people to their side, and provide leverage over opponents. The manner and procedure in which the fuel subsidy removal policy was conducted shows support for a political framework. This of course, the generality of the people did not agree with. According to CPPA (2012) there is evidence that the countries that have successful implemented fuel subsidy removal took a phased or gradual approach, engaged in conscientious research prior to implementation and followed a rigorous approach to policy making. In addition, effective communications and a fair level of trust between citizens and government are other critical success factors in the policy of subsidy removal.

The purpose of marketing research which is a crucial aspect of marketing in public policy decision making is to assists decision makers to understand the environment and the public needs. It helps to remove some of the uncertainty by providing relevant information about the environment and the public. It provides sound information which is not based on gut feeling, intuition, or even pure judgment. This information enhances the effectiveness of decisions made by public officials. According to Alan (2006) marketing research involves the following steps:

- The collection of information using a wide range of sources and techniques. Such information may be acquired from published sources, observing behaviour or through direct communication with the people being researched.
- The analysis of information. The information needs to be analyzed, developed and applied if it is to be actionable and relevant to the situation.

- The communication and dissemination of information. The effective presentation of information transfers understanding of its content and implications to a wider audience of relevant decision makers and interested parties.

In marketing, there is nothing wrong with increasing the price of a product, but it will be a disastrous decision if the firm concerned is not able to provide an acceptable justification for the increase because the customers will switch to other products if there are substitutes or resist vehemently if it is a monopoly. However, if the consumers are convinced that the increment will provide additional value, then of course, they will still patronize the product. In the absence of relevant information, public response to governmental policies cannot be predicted reliably or accurately. Nigerians have constantly been failed and neglected by previous governments to the extent that any policy that does not have immediate benefit is often viewed as a deceptive policy. The average Nigerian is skeptical to government policies whether good or bad because in the eyes of the masses, they are all the same with different faces and names (National Impact, 2012).

Fuel subsidy removal is usually tagged an International Monetary Fund (IMF) policy. According to Agbon (2012) the IMF document provided some guidance before the implementation of fuel subsidy removal. "They advised on how to identify political opponents of the fuel subsidy removal program, how to do a publicity campaign, how to set up a program aimed at using the money generated, how to time the subsidy removal, how to make promises of transport buses, education, health, roads and give money to the poor if necessary". This indicates that for effective implementation of the policy as it is an area that is largely affecting the lives of most Nigerians, several questions and issues needed to be addressed and marketing research is all about providing answers to strategic questions. To ensure the success of the policy the following questions are critical and the answers are necessary:

- Who are the opponents of the fuel subsidy removal program?
- What are the justifications for the policy and how do we convince the opponents that the policy is the best option at the moment?
- What is the most appropriate medium to use to sensitize the opponents?
- How do we set up a program aimed at using the money generated?
- What is the right time to remove the subsidy?
- Are there infrastructural facilities that can ameliorate the effect of the policy?
- How do we identify and assist those that will be adversely affected by the policy?

Bridgetown (2007) also examined the elements to consider in public policy decision making. He listed the following:
1. Goals of the Policy: What are the goals? Will it eliminate the problem? Will it alleviate the problem? Will it prevent the problem from worsening?
2. Causal Model Underlying Public Policy: What is the causal theory supporting the policy? If the policy is implemented, do we know all the consequences? How can we determine causality (if possible)?
3. Tools of the Policy: What instruments will be used to implement the policy? What is the degree of coercion? Will the tools rely on incentives, persuasion, information and capacity -building?
4. Targets of Policy: Whose behavior is supposed to change? Are there direct and indirect beneficiaries? What assumptions about the target population underlie the choice of tools?
5. Implementation of the Policy: How will the policy be implemented? Who will define the criteria for implementation? Who will enforce the implementation? And what is the best time for implementation?

If these issues were properly addressed, it would have gone a long way to reduce if not bring to a halt the kind of protests that engulfed the country immediately the policy was announced. Governments that have rushed subsidy reforms without preparing the population for the changes, and without providing targeted support to particularly disadvantaged groups, have often had to reverse the policy in the face of widespread opposition (World Bank, 2010; CPPA, 2012). Countries like Niger and Ghana had taken a similar policy measure to their economic advantage without any form of protest because some of the issues stated above were addressed before the policy took effect (Agbon, 2012). Although the federal government tried to convince some Nigerians especially the state governors that the policy measure was capable of enhancing the economic fortunes of this country, it is evident that the consultations were not actually to seek the input of the people, or to gauge their feelings and opinions concerning the issue, instead they were meant to inform them of their resolve to go ahead with the policy (Onyishi, et al. 2012). So, the federal government should have extended the period of awareness sensitization to enable all stakeholders and not just a selected few to actually appreciate government's good intentions.

Aliu (2011) noted that the former president of the National Union of Petroleum and Natural Gas Workers (NUPENG) Peter Akpattasson stated that the timing for the planned removal of the subsidy was wrong because the right investment climate in the sector that would help regulate pricing has not been created. What is important at this point in time is to start working towards achieving such economic environment where the level of manipulation will be curtailed and investors will be encouraged to come in and set up refineries. Moreover, there was a subsisting understanding between labour and the federal government in 2009 that removal of subsidy will not begin, until certain conditions have been met. These conditions which include fixing of the refineries and building of new ones, regular power supply and provision of other social infrastructure, such as railways and repairs of roads as well as the elimination of the corruption associated with supply and distribution of petroleum products in the downstream sector of the oil industry have not been met.

Based on these circumstances, it could be inferred that the government failed to establish a structure for a sustainable fuel subsidy removal. There ought to be existing structures to address the high cost of living and cost of food items. This should also address the problem of poor infrastructure such as lack of electricity, good roads, lack of potable water, shortage of quality habitable accommodations. The views of the public are that government must fast track the turnaround maintenance of the four refineries and encourages the building of new ones. This will help reduce the dependence on importation of refined products and protect the economy from the volatility of global oil prices. Also, the power sector and its problems and other utilities must be properly addressed. The governance structure should be more cost effective and corruption must be more effectively tackled. There is a seeming agreement among Nigerians that the Nigerian National Petroleum Corporation (NNPC) is corrupt and needs a complete reorganization and persons found to be guilty should be appropriately punished. The allowances of members of the cabinet including the national assembly members should be drastically reduced and their activities checked. It is also imperative that the estimated US$6B savings per annum be used judiciously. If Nigerians can see where, how and when the US$6B savings is utilized then Nigerians may bear the pain.

There is also the need to focus marketing on public policy issues. As noted by Dyer and Shimp (1977) marketing is somehow alienated from public policy. They pointed out that marketers should also take the initiative in doing research on public policy issues and bring it to the attention of the policy makers. Disciplines in the social sciences such as economics, sociology, political science, public administration are usually more concerned about problems of governmental administration, management and operations. They should more than ever before focus their researches on how public officials can make effective decisions. This is necessary because effective research could significantly improve many public policy decisions and help to carry the people along. In addition, if research is to have an impact on public policy, it must be useful and available to the policy maker. Studies that are relevant to public policy decision making should be made accessible to policy makers. Such research should be practical, unambiguous and devoid of suspicion.

## VII. Conclusion

The principles and concepts of marketing have been found to be relevant not only to business organizations but also to the society. Any government that satisfies the needs and aspirations of the public will always be recognized, applauded, supported and re-elected as we have observed in some states in Nigeria and in other countries. For government to understand what the public want and how they want it, there is need for marketing research. The purpose is to provide decision makers with relevant, accurate, reliable, valid, and current information. Such sound research information will enable public policy makers to better understand the public that they are trying to influence and to adequately educate the public thereby improving public policy decisions. Marketing research is objective and could significantly improve many public policy decisions. It attempts to provide accurate information that reflects the true state of affairs. However, it should be conducted impartially. It should be free from the personal or political biases of the decision makers because research which is motivated by personal or political gain involves a breach of professional standards.

## References

[I]     Agbon, I. (2012). IMF and Fuel Subsidy Removal in Nigeria, Retrieved from: http://www.seunfakeze.wordpress.com/... /imf-and-fuel-subsidy-removal-in-nigeria.
[2]     Ajaero, C. (2012). Nigeria Grounded as Protests over Fuel Subsidy Removal Cripple Activities      Nationwide, Newswatch, *Vol.55. No.2,* Jan 23, 14-23.
[3]     Akanbi, F. and Agbo, M. (2012). Strike: Nigerians Count their Losses, *This Day,* Sunday, January 15, 24.
[4]     Alan, W. (2006). *Marketing Research: An Integrated Approach,* Harlow: Prentice Hall.
[5]     Alexander, C. (2012). Fuel Subsidy Removal: Repositioning Nigeria Economy, Businessday, January 9, *Retrieved from:* http://www. Businessday online.com.
[6]     Aliu, A.O. (2011). Former NUPENG president opposes proposed policy, *Guardian*, Tuesday, October11, 48.
[7]     Balouga, J. (2012). The Political Economy of Oil Subsidy in Nigeria, *International Association for Energy Economics*, *Second Quarter*, 31-35.
[8]     Beri, G.C. (2008). *Marketing Research*, New Delhi: McGraw-Hill Companies.

[9]   Brent, J.R. and LaBrèque, R.J. (1975). Marketing Research and Public Policy: A Functional Perspective, Journal of *Marketing*, *Vol 39, No. 3*,12-19.
[10]  Bridgetown, B. (2007). Policy Analysis and Decision-Making, *Strategic Health Development Area, PAHO/WHO*
[11]  Centre for Public Policy Alternatives (CPPA, 2012). Nigeria: Fuel Subsidy, A Desktop Study, Retrieved from: http://toluogunlesi.files. wordpress.com/2012/01/fuel-subsidy-desktop-study-report.copy 1.
[12]  Cooper, D.R. and Schindler, P.S. (2006). *Marketing Research,* Irwin: McGraw-Hill.
[13]  Dyer, R.F. and Shimp, T.A. (1977). Enhancing the Role of Marketing Research in Public Policy Decision Making, *Journal of Marketing*, *Vol 41, No. 1*,63-67.
[14]  Ering, S.O. and Akpan, F.U. (2012). The Politics of Fuel Subsidy, Populist Resistance and its Socio-Economic Implications for Nigeria, *Global Journal of Human Social Science*, *Vol 12 Issue 7 Version 1.0 April*, 13-20.
[15]  Gundlach, G.T. (2007). The American Marketing Association's 2004 Definition of Marketing:Perspectives on Its Implications for Scholarship and the Role and Responsibility of Marketing in Society, *Journal of Public Policy and Marketing*, Vol. 26 (2), 243–250
[16]  Hogan, T.D. (2005). *Data for Effective Policy and Decision-Making in Indiana: Assessing its Availability, Accessibility, and Analysis,* India: Central Indiana Corporate Partnership.
[17]  Kotler, P. and Keller, K.L. (2006). *Marketing Management*, England: Pearson Education Inc.
[18]  Lusch, R.F. (2007). Marketing's Evolving Identity: Defining our Future, *Journal of Public Policy and Marketing*, Vol 26 (2), 261-268.
[19]  NationalImpact(2012).Tackle Corruption not Just Fuel Subsidy, *Retrieved from: http://www.nationalimpact.net/2012/02/tackle - corruption-not-just-fuel-subsidy/.*
[20]  Nwadialo, U. (2012). Fuel subsidy removal: A Nigerian dilemma, *Vanguard, January 9, Retrievedfrom: www.vanguardngr.com /.../fuel -subsidy-removal-a-nigerian-dilemma/.*
[21]  Nzeagwu, U. and Gyamfi, C.C. (2011). NANS Cautions against Fuel Price Increase, Cleric Decries Poverty Level, *Guardian*, Tuesday, September 27, 8.
[22]  Okonmah, T.N. (2012). Fuel Subsidy Removal: Not the Solution to Nigeria's Economic Recovery, *Vanguard, January 9, Retrieved from: www.vanguardngr.com/.../fuel-subsidy-removal-not-the-solution-to-...*
[23]  Onyishi,A.O., Eme, O.I. and Emeh, I.E. (2012). The Domestic and International Implications of Fuel Subsidy Removal Crisis in Nigeria, *Kuwait Chapter of Arabian Journal of Business and Management Review, Vol. 1, No.6; February*, 57-80.
[24]. Ovaga, O.H. (2012). Subsidy in the Downstream Oil Sector and the Fate of the Masses in Nigeria, *Kuwait Chapter of Arabian Journal of Business and Management Review Vol.1, No.6; February*, 15-34.
[25]  Wilkie, W.L. and Gardner, D.M. (1974).The Role of Marketing Research in Public Policy Decision Making, *Journal of Marketing*, *Vol 38, No. 1*, 38-47.
[26]  World Bank (2010). Subsidies in the energy sector: An Overview, *The World Bank Group*, *Retrieved from: http://siteresources.worldbank.org/EXTESC/Resources/Subsidy_background_paper.pdf.*